\documentclass[twocolumn,showpacs,preprintnumbers]{revtex4}
\usepackage{amssymb}
\usepackage{amsmath}
\usepackage{graphicx}
\usepackage{epsfig}

\begin{document}
\title{\bf  Point-contact spectroscopy of  Al- and C-doped
MgB$_2$. Superconducting energy gaps and scattering studies.}

\author{P. Szab\'o,$^{1}$ P.  Samuely,$^{1}$ Z. Hol'anov\'a,$^{1}$ 
S. Bud'ko,$^{2}$ P. C. Canfield,$^{2}$ and J. Marcus$^{3}$}

\address{$^1$Centre of  Low Temperature Physics, IEP Slovak
Academy of Sciences\& P.J.\v Saf\'arik University, Watsonova
47, SK-04001 Ko\v sice, Slovakia\\
$^2$Ames Laboratory and Iowa State University, Ames, IA 50011, USA\\
$^3$LEPES CNRS, F-38042 Grenoble Cedex 9, France.
}

\date{\today}

\begin{abstract}

The two-band/two-gap superconductivity in aluminium and carbon doped MgB$_2$ has been addressed by the point-contact spectroscopy. Two gaps are preserved in all samples with $T_c's$ down to 22 K. The evolution of  two gaps as a function of the critical temperature in the doped systems suggest the dominance of the band-filling effects but for the increased  Al-doping the enhanced interband scattering approaching two gaps must be considered. 
The magnetic field dependences of the Andreev reflection excess currents as well as zero-energy density of states determined from the experimental data are used to analyze the intraband scattering.  It is shown, that while the C-doping increases the intraband scattering in the $\pi$-band more rapidly then in the $\sigma$ band, the Al-doping does not change their relative weight. 

\end{abstract}

\maketitle




\section{INTRODUCTION}
Magnesium  diboride owes  its high  critical temperatures of
40 K  \cite{akimitsu} to the interplay  between two distinct
electronic bands crossing the Fermi level. About half of the
quasiparticles  belongs  to  the  quasi two-dimensional hole
$\sigma$  band  and  the  Cooper  pairs  there  are strongly
coupled via the boron vibrational mode $E_{2g}$. The rest of
the  quasiparticles resides  in the  three-dimensional $\pi$
band  with  a  rather  moderate electron-phonon interaction.
Without  the effect  from the  $\sigma$-band the  transition
temperature $T_c$ would be just a fraction of those found in
the   real    MgB$_2$   \cite{liu,choi}.   The    two   band
superconductivity is the most  spectacularly revealed by the
presence of two very distinguished energy gaps, large in the
$\sigma$  and  small  in  the  $\pi$  bands \cite{szabo,giubileo,iavarone,suderow,tsuda}.  
Superconducting properties of such a system 
are  sensitive to the scattering. 
Due to coexistence of charge carriers from two almost completely separated bands two {\it intraband} and one {\it interband} scattering channels have to be distinguished.
The {\it  interband} scattering  by non  magnetic scatterers  is
supposed   to have   a particularly  strong   effect   in  a  two-band
superconductor: it  will blend  the strongly
and  weakly  coupled  quasiparticles,  merge  two  gaps  and
consequently decrease  $T_c$. In the  case of MgB$_2$  $T_c$
can  drop  down  to  about  20  K  \cite{choi}. Fortunately,
a different symmetry of the bands ensures that the interband
scattering remains small also  in very dirty MgB$_2$ samples
which  show  about  the  same  $T_c$  as the purest material
\cite{mazin}. A systematic decrease  of $T_c$ is achieved in
substituted MgB$_2$ samples. The  only on site substitutions
by non magnetic  elements which are known so  far are carbon
for  boron  and  aluminium  for  magnesium.  The  carbon and
aluminium atoms in MgB$_2$ take  indeed a role of scatterers
but they also dope the  system with one extra electron which
inevitably leads to the filling  effect in the $\sigma$
band, with strongly coupled holes.  Recently,  Kortus  {\it  et  al.}  \cite{kortus} have
introduced  a  model  incorporating  both effects: interband
scattering and  band filling in  MgB$_2$. The former  effect
leads  to an  increase of  the small  gap $\Delta_{\pi}$ and
decrease  of the  large $\Delta_{\sigma}$  while the  latter
suppresses both $\Delta_{\pi}$  and $\Delta_{\sigma}$. There
is a  rather controversial situation as  far as the strength
of both effects  in the Al and C-doped  MgB$_2$ is concerned
\cite{kortus,samuelyk}.  

In contrast to the interband scattering an increase of the scattering within the bands 
does not have any effect on the two gaps. But, the selective tuning of the {\it 
intraband} scattering can lead to the expressive variation of the 
values of the upper critical magnetic field $H_{c2}(0)$ and its anisotropy \cite{angst,gurevich}. 
Upon increased  C-doping $H_{c2}(0)$ is increased significantly in both principal
crystallographic directions (parallel  to the $ab$ hexagonal
boron   layers    and  the   $c$-axis    direction). On the other hand  Al
substitution  suppresses  $H_{c2}$(0)  in  the $ab$-plane
direction  and  in  the  $c$-axis  direction  $H_{c2}$  is  hardly
changed. This difference behavior is due to a different influence of C- and Al-doping on intraband scatterings.
While the strong increase of $H_{c2}(0)$ with the C-doping is  due to graded dirty limit conditions,  Al-doped samples stays still in the clean limit and the decrease  of $H_{c2}$  is just  a   consequence  of  the  lower $T_c$ \cite{angst}. A weight of scatterings in separated bands is still not clear. 

In this paper we present the systematic study    of    the    superconducting    energy    gaps   in
both Al- and C-doped magnesium diborid systems at stoichiometries 
Mg$_{1-x}$Al$_x$B$_2$   with   $x=0,   0.1$   and   0.2  and Mg(B$_{1-y}$C$_y$)$_2$  with  $y=0,   0.021,  0.038,  0.055, 0.065$  and  0.1.  For  the   carbon  doping  this  work  is a supplement  to  our  previous  studies  \cite{holanova} on Mg(B$_{1-y}$C$_y$)$_2$ with $y=0, 0.021,  0.038,$ and 0.1. 
The influence of both substitutions for the scattering processes is discussed.

\section{EXPERIMENT}

Polycrystalline  samples  of   Al-doped  MgB$_2$  have  been
prepared  by a  two step  synthesis at  high temperatures
as       Mg$_{0.8}$Al$_{0.2}$B$_2$        and
Mg$_{0.9}$Al$_{0.1}$B$_2$ with  $T_c$ = 30.5  K and 23.5  K,
resp.   \cite{angst}.   The   carbon   substituted   samples
Mg(B$_{(1-y)}$C$_y$)$_2$ with  $y = 0, 0.055,  0.065, 0.1$ and
with the respective transitions at $T_c$  = 39 K, 33 K, 28.2
K,  and  22  K,  were  synthesized  in  the  form of pellets
following the  procedure described in  Ref.\cite{angst} from
magnesium lumps and B$_4$C powder. For $y = 0, 0.021, 0.038$
we have worked with the wire segments with $T_c$ = 39 K, 37.5 K,
and  36.2 K \cite{wilke}. The undoped MgB$_2$ sample with $T_c = 39$ K in a dirty limit with the high upper critial field has been prepared  from boron and magnesium  powder  as described elsewhere \cite{szabo}
   
The point-contact (PC)
measurements  have been  realized via  the standard  lock-in
technique in a special point-contact approaching system with
lateral and vertical movements of the PC tip by a differential
screw  mechanism.  The  microconstrictions  were prepared in
situ  by  pressing  different   metallic  tips  (copper  and
platinum  formed either  mechanically or  by electrochemical
etching) on different parts  of the freshly polished surface
of  the  superconductor.  Point-contact  measurements on the
wire segments have been
performed in a "reversed" configuration - with the
Mg(B$_{1-x}$C$_x$)$_2$ wire as a  tip touching softly a bulk
piece of electrochemically cleaned copper.
$T_c$'s in both series of substitutions have been determined
from the resistive transitions
and   also  from   the  temperature   dependences  of   the
point-contact spectra. The  transition temperatures found by
both methods were essentially  the same, indicating that the
information    obtained   from   a more   surface    sensitive
point-contact technique is relevant also for the bulk.

The point-contact    spectrum   measured    on   the    ballistic
microconstriction   between   a   normal   metal   (N)   and
a superconductor (S) consist of  the Andreev reflection (AR)
contribution and the  tunneling contribution \cite{btk}. The
charge    transfer    through    a    barrierless   metallic
point-contact  is  realized  via  the  Andreev reflection of
carriers. Consequently  at $T = $0  the PC current as  well as
the PC  conductance inside the gap  voltage ($V< \Delta /e$)
is  twice  higher  than  the  respective  values  at  higher
energies  ($V>>\Delta /e$).  The presence  of the  tunneling
barrier  reduces the  conductance at  the zero  bias and two
symmetrically  located peaks  rise  at  the gap  energy. The
evolution  of  the  point-contact  spectra  between the pure
Andreev reflection  and the Giaver-like  tunnelling case has
been       theoretically        described       by       the
Blonder-Tinkham-Klapwijk   (BTK)   theory   \cite{btk}.  The
point-contact  conductance data  can be  compared with  this
theory using  as input parameters  the energy gap  $\Delta$,
the parameter $z$ (measure for the strength of the interface
barrier)   and  a   parameter  $\Gamma$   for  the  spectral
broadening \cite{plecenik}. In  any case the  voltage dependence of  the N/S
point-contact   conductance   gives   direct   spectroscopic
information   on   the   superconducting   order   parameter
$\Delta$.  For   the  two-gap  MgB$_2$   superconductor  the
point-contact conductance $G$ can be expressed as a weighted
sum of  two partial BTK  conductances: those from  the quasi
two-dimensional  $\sigma$-band  (with  a  large  gap $\Delta
_{\sigma}$) and  those from the 3D  $\pi$-band (with a small
gap $\Delta_{\pi}$)
\begin{equation}
G  =  \alpha  G_{\pi}  +  (1-\alpha )G_{\sigma}.
\end{equation}
The  weight  factor  $\alpha$  for  the  $ \pi$ - band 
contribution can vary from 0.6 for the point-contact current
strictly  in  the  MgB$_2$  $ab$-plane  to  0.99 of $c$-axis
direction \cite{brinkman}.

\section{RESULTS AND DISCUSSION}

\subsection{Zero field point-contact spectroscopy}

A large number  of the point-contact  measurements have been
performed  on the  tested samples  of pure  MgB$_2$ and both
substituted  series. The  point contacts  revealed different
barrier   transparencies   changing   from   more   metallic
interfaces to intermediate  barriers with $0.4<z<0.8$ for
the aluminium  doped serie and  $0.4<z<1.2$ for the  C-doped
samples. By trial and error we looked for the spectra showing both gaps of a polycrystalline specimen.
For  more   detailed  studies  we   chose  those  junctions
revealing  the  spectra  with   a  low  spectral  broadening
$\Gamma$, i.e. with an intensive signal in the normalized PC
conductance. As  a result we  succeeded to obtain many point
contact spectra with well  resolved two superconducting gaps
directly  in the  row data  for all  of the  studied substitutions
except  for the  highest dopings,  Mg$_{0.8}$Al$_{0.2}$B$_2$
and   Mg(B$_{0.9}$C$_{0.1}$)$_2$.  Figure   1 displays  the
representative spectra of the aluminium doped serie and Fig.
2 resumes  the results  for  the  carbon doped  MgB$_2$. The
experimental data are presented by the full lines, while the
fits  to  the  two-gap  formula  are  indicated  by the open
circles. All
shown  point-contact  spectra  have  been  normalized to the
spectra measured in the normal  state at a temperature above
$T_c$.  The upper  curves are  shifted for  the clarity. The
retention of two gaps for  all dopings apart from the highest
substitution is evident.

For the
highest  dopings   (always  with   a  larger   spectral
broadening, $\Gamma \approx 0.2\Delta$),  the spectra in Figs. 1
and 2 reveal
only one pair of peaks without an apparent shoulder at the
expected  position  of  the   second  gap.  For  those  data
measured at 1.6 K (the rest  of tha data was measured at 4.2
K) also the  one-gap fit is presented. The  one-gap fit can
reproduce  either the  height of  the peaks  or the width of
them, while the two-gap fit  can reproduce both. The size of
the  apparent gap is  well indicated  by the  peak position, which is
about 2 meV for  Mg$_{0.8}$Al$_{0.2}$B$_2$ and 1.6 meV for 
Mg(B$_{0.9}$C$_{0.1}$)$_2$.
But, this size is too small to explain the superconductivity with
the respective $T_c$'s  equal to 23.5 K and  22 K within the
single-gap BCS scenario suggesting, that we are dealing with a smeared two-gap structure.
In the case of the 10\% C-doped MgB$_2$ an existence of the
large gap was clearly shown by the specific heat mesurements
\cite{ribeiro}. Moreover, an
existence  of the  both gaps  was strongly evidenced  in a  single
experiment   by  our   previous  point-contact  spectroscopy
measurements in applied magnetic field \cite{holanova}. There, the spectrum at $H$ = 0.7 T showed that the contribution from the $\pi$ band was
partially  suppressed  and  the  large-gap  shoulder clearly
appeared near  3 mV besides the gap peak at 1.6 mV. A  similar effect has
the magnetic field in the case of 20 \% Al doped MgB$_2$:
As shown in the top panel of Fig. 5, the peak position being at
about 2 mV in zero field, is shifted towards 3 mV at 1 Tesla.
In the  case of a single-gap  superconductor the application
of  magnetic  field  can  only  lead  to  a shrinkage of the
distance  between  the  peak  positions  in the point-contact
spectrum. It is a simple  consequence of an introduction of
the vortices  and magnetic pair-breaking  \cite{sam385}. 
The shift of the peak of the PC spectra to higher voltages 
demonstrated  in  Fig.  5  is then due to
an interplay  between the two  gaps, because  the dominance of the $\Delta_{\pi}$ peak at zero field  is suppressed in increasing field and $\Delta_{\sigma}$ contributes more. Thus, the  retention of
two gaps is proved in  the  all  presented doped
MgB$_2$ samples with $T_c$'s from 39 down to 22 K.
\begin{figure}[t]
\includegraphics[width=7.6 cm]{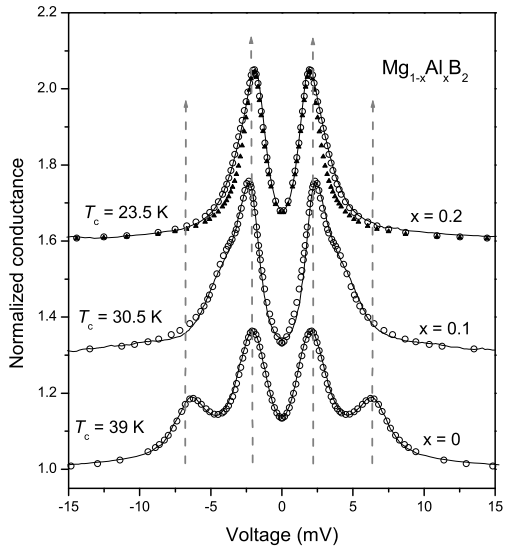}
\caption{Normalized point-contact spectra on Mg$_{1-x}$Al$_x$B$_2$ with $x = $ 0, 0.1 and 0.2 - solid lines. Symbols show the fitting curves: open circles for the two-gap BTK model and solid triangles for the one-gap BTK model. The spectra are vertically shifted for the clarity. The dashed vertical arrows emphasizes the tendency of the evolution of the gaps.}
\label{fig:fig1}
\end{figure}

The statistics  of  the   energy gaps  obtained from
fitting to the  two-band BTK formula is shown  in Fig. 3,
for  each  C  and  Al concentration.  The left coordinates
indicate the transition temperature of the junctions, while
the right one  counts the number of junctions  with the same
size  of the  gaps. The  energy width  of a particular count
indicates  the  fitting  uncertainty.  All  the samples, the
undoped MgB$_2$ as well  as the
aluminium and
carbon doped  material reveal a  certain distribution of  the
small  and  large  gaps  but  the  two  gaps are well
distinguishable  in the histogram and  no   overlap  of  $\Delta_{\sigma}$  and
$\Delta_{\pi}$  is  observed.  Already  from  this figure a
general tendency  is evident, namely  that the reduction  of
the  large  gap  is  proportional  to the decreasing transition
temperature. On the other hand  the changes in the small gap
are proportionally smaller.
\begin{figure}[tp]
\includegraphics[width=8.2 cm]{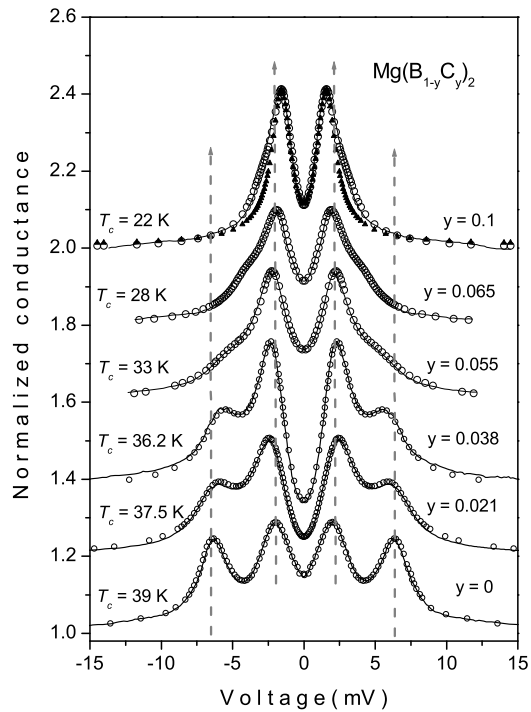}
\caption{ Normalized point-contact spectra measured on Mg(B$_{(1-y)}$C$_y$)$_2$ with $y = $ 0, 0.021, 0.038, 0.055,  0.065 and 0.1 - solid lines. Symbols show the fitting curves: open circles for the two-gap BTK model and solid triangles for the one-gap BTK model. The spectra are vertically shifted for the clarity. The dashed vertical arrows emphasizes the tendency of the evolution of the gaps.}
\label{fig:fig2}
\end{figure}

Figure 4 displays  the energy gaps as a function of $T_c$'s.
The  points (open squares for Al dopings and solid circles for 
C ones)  are  positioned  at  the averaged  energies of  the gap  distributions from Fig. 3 and  the error
bars  represent the standard deviations. The large gap 
$\Delta_{\sigma}$ is essentially decreased linearly with the respective $T_c$'s.  The behavior 
of $\Delta_{\pi}$'s is more complicated. For both kind of substitutions
 the gap is almost unchanged at smaller dopings. In the case of C-doped 
MgB$_2$ it holds   clearly down to $T_c = $ 33 K. In (Mg,Al)B$_2$
$\Delta_{\pi}$ of the 10 \% Al doped sample seems to even slightly increase. 
But for the highest dopings $\Delta_{\pi}$ decreases in both doping cases. The 
dashed lines in Fig. 4
are calculations of Kortus {\it et al.} \cite{kortus} for the case of a pure 
band filling effect and no interband scattering. Although, the constant 
$\Delta_{\pi}$ is not reproduced,  the $T_c$ dependence of $\Delta$'s
for carbon doping are broadly well described. The solid lines show the Kortus's 
calculations including the interband scattering with the rate $\gamma_{IB} = 
1000x$cm$^{-1}$ (or $2000y$cm$^{-1}$).
As can be seen the gaps in the Al doped samples cannot be accounted without 
an interband scattering but it is  smaller than in the presented  
calculations.

\begin{figure}[tp]
\includegraphics[width= 8.2 cm ]{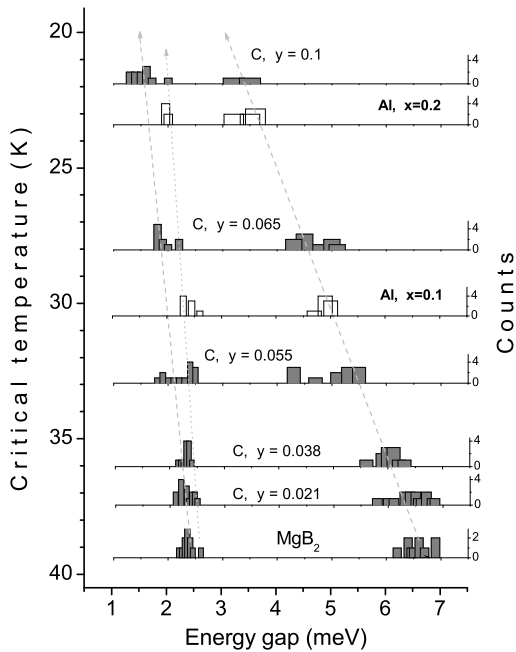}
\caption{Distribution of the superconducting energy gaps of Mg(B$_{(1-y)}$C$_y$)$_2$ - gray columns and Mg$_{1-x}$Al$_x$B$_2$ - open columns. The dashed and dotted lines are guides for the eyes.}
\label{fig:fig3}
\end{figure}

\begin{figure}[tp]
\includegraphics[width= 8 cm ]{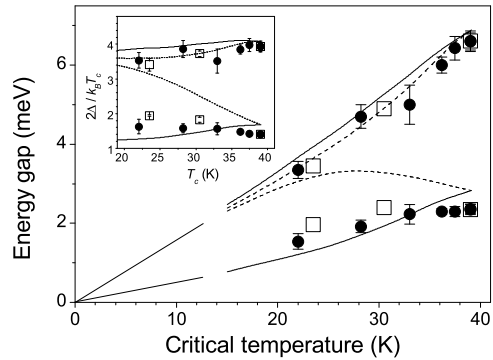}
\caption{Averaged values of the superconducting energy gaps  as a function of the critical temperatures of Mg(B$_{(1-y)}$C$_y$)$_2$ - solid circles and Mg$_{1-x}$Al$_x$B$_2$ samples - open squares. The error bars represents the standard deviations in the distribution shown in Figure 3. The lines - see the text.}
\label{fig:fig4}
\end{figure}

Recently, similar results on the superconducting energy gaps have been obtained on the Al-doped single crystals  
by some of us and other authors \cite{klein06}  exploring the specific 
heat as well as the point contact spectroscopy measurements. Those data also show a 
tendency of approaching of the two gaps with their possible merging 
in the Al-doped samples with $T_c$ below 10-15 K. 
It is worth noticing that those and here presented results do not show 
any decrease of the large 
gap below the canonical BCS value for the Al-doped MgB$_2$ with $T_c$'s 
below 30 K found earlier \cite{putti1}.

For the carbon doped samples we have not found any stronger 
tendency to blend both gaps
down to $T_c = 22 $K, being in contrast with the data obtained 
on single crystals 
by Gonnelli {\it et al.} \cite{gonnelli}.
On the other hand our data are very compatible with the recent ARPES 
measurements of 
Tsuda {\it et al.} where both gaps have been  directly seen in the row data down 
to $T_c$ of 23 K.
A presence of two gaps is also evidenced by the point contact measurements of 
Schmidt {\it et al.} 
\cite{schmidt} on the heavily carbon doped MgB$_2$. We  remark  that  the data showing no merging of 
the gaps on the heavily doped MgB$_2$ with $T_c$ down to 22 K  have been collected from the measurements on the samples of different 
forms (wires, sintered pellets,
polycrystals) prepared by different methods. Moreover 
it  is  very  important  that two gaps have been directly
experimentally evidenced in the row data without any dependece on the model or 
fit.

The theoretical calculations of Erwin and Mazin \cite{erwin} on merging 
of the gaps
due to a substitution are also supporting the picture presented here. By them 
the carbon 
substitution on the boron site should have zero effect on the merging of the 
gaps.
It is due to the fact that a replacing boron by carbon does not change the local 
point symmetry
in the $\pi$ and $\sigma$ orbitals which are both centered at the boron sites. 
Then, the 
interband scattering is of little probability without incorporating extra 
defects producing a 
change of the symmetry of orbitals. 
On the other hand much bigger effect 
is expected for the out of plane substitutions  (Al instead of Mg) or defects which would indeed 
change the 
local point symmetry. The substitution of Al instead of Mg leads also to a 
significant
decrease of the $c$-lattice parameter \cite{puttic} (it is basically nullite in 
case of 
carbon doping), which helps the interlayer hopping from a $p_z$ orbital ($\pi$-band) in one 
atomic layer 
to a $\sigma$ bond orbital in the next one. Of coarse, a particular strength of 
the 
interband scattering can scatter among samples due to deffect and thus the 
merging of two gaps 
could be sample dependent.

\subsection{Magnetic-field effects}

\begin{figure}[tp]
\hspace{-5mm}
\includegraphics[width= 7 cm ]{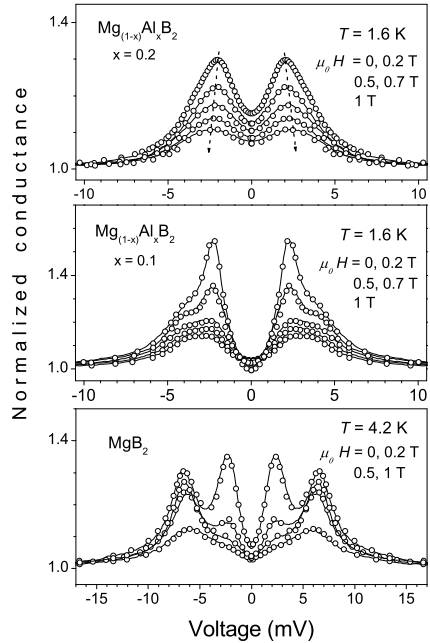}
\caption{Influence of the applied magnetic field on the point-contact spectra  of Mg$_{1-x}$Al$_x$B$_2$ - solid lines. Open circles show fitting results for the two-band mixed state BTK formula.}
\label{fig:fig5}
\end{figure}
In the following we present  the effect of magnetic field on
the point contact spectra of the aluminium as well as carbon
doped MgB$_2$ samples. The field effect has been measured on
a number  of  samples  of  each  substitution.  The field was
always  oriented perpendicularly  to the  sample surface and
parallel with the  tip having the point contact  area in the
vortex state.

Figure 5 displays the  typical magnetic field dependences of
the PC  spectra (lines) obtained  on Mg$_{(1-x)}$Al$_x$B$_2$
samples with x = 0, 0.1  and 0.2. The zero-field spectrum of
the  undoped MgB$_2$  (bottom panel)  reveals an  expressive
two-gap   structure.  At  small   fields  the  peak
corresponding  to  the  small  gap  $\Delta_{\pi}$  is  very
rapidly  suppressed and  only the  peak responsible  for the
large  gap $\Delta_{\sigma}$  is apparently  visible above 1 Tesla. At 10 \% of Al doping the zero field
spectra  also  show  both  gaps  but  the  large  one is now
revealed  more as  a pronounced  shoulder. Again  at a small
magnetic field the $\pi$-band-gap peak is rapidly suppressed
but its signature is still  distinguishable at 0.5 Tesla. At
higher fields the signatures of both gaps are not resolvable
due to an interference of the related peaks placed at closed
positions.   The  resulting   maximum  is   located  in  the
intermedium  position between  them. At  the highest  doping
(20  $\%$ Al)  such an  interference of  even closer gaps is
graduated,  so already  at zero  magnetic field  one can see
only one pair of peaks at  the position of the small gap (it
has  always much  higher weight  at zero  field) and a minor
shoulder at the voltage of  the large one. In the increasing
field  the peak  is broadened  and its  position shifted  to
higher  voltages  (indicated  by  arrows).  As mentioned above the  shift is an
indication of  two gaps present  in the spectrum with 
an increasing  weight of the large-gap peak with
respect to the small one.

The in-field  measurements have also  been performed on  the
carbon doped samples. In  our previous paper \cite{holanova} they
have  been   demonstrated  on  the  heavily   doped,  10  \% C MgB$_2$ sample.  

In  two sets  of measurements  and subsequent analysis  on the aluminum and
carbon doped samples two  different types of undoped MgB$_2$
samples  have been  used.  The  carbon doping  increases the
upper  critical  fields  \cite{angst}  despite the decreased
$T_c$ and both superconducting gaps. Thus, these samples are
driven to the dirty limit by  doping. That is why we compare
the  6.5  \%  and  10  \%  C-doped  samples with the undoped
MgB$_2$ crystal  being already in  a dirty limit  ($T_c \approx
39$ K, $H_{c2||c} \approx 5$  T) \cite{szabo}. On the other
hand  the  aluminum  doping  leads  only  to  a  decrease of
$H_{c2}$ for the field in $ab$ planes or a constant value in
perpendicular    orientation    with    $H_{c2||c}    =   3$
T \cite{angst}, so  we have chosen  the pure MgB$_2$  sample
with  $T_c \approx  39$ K  and $H_{c2||c}  \approx 3$  T as
a reference.

From the in-field spectra the excess current $I_{exc}$ has
been  determined. It  was calculated  by integrating  the PC
spectra after subtracting an area below unity. When a single
gap superconductor is in the  vortex state, each vortex core
represents  a  normal  state  region  and this region/vortex
density increases linearly up to the upper critical magnetic
field.  Thus, the  PC excess  current ($\propto \Delta$) is
reduced linearly in increasing  magnetic field by the factor
of $(1-n)$, where $n$ represents a normal state region - the
PC  junction area  covered by  the vortex  cores. The  field
dependences of the excess currents $I_{exc}$ obtained on the
carbon and aluminum  doped samples are shown in  Fig. 6a and
6b, respectively.  The magnetic field  coordinates have been
normalized to  an appropriate $H_{c2||c}$ referred  to as $h
= H  / H_{c2||c}$ \cite{footnote}.  In a  none case  the linear  decrease of
$I_{exc}$  was  observed.  On the contrary,  a strong non-linearity
observed in  all curves approves  a presence of  two gaps in
the quasiparticle  spectrum of the  material. There, a  more
rapid  fall  of  $I_{exc}(h)$  at  low  magnetic  fields  is
ascribed to a strong filling of the $\pi$-band gap states up
to a  virtual/crossover  $\pi$-band upper critical  field
$H_{c2,\pi}$. Then,  at higher fields,  the low temperature superconductivity
is   maintained   mainly   by  the
$\sigma$-band  \cite{kogan}, with  the $\sigma$  gap getting
filled smoothly up  to the real upper critical  field of the
material $H_{c2} = H_{c2,\sigma}$. Due to a small, but finite coupling of
the two  bands, the superconductivity  is maintained also  in
the $\pi$-band above $H_{c2,\pi}$, but,  as shown e.g. in the
tunneling data  of Eskildsen {\it  et al.} \cite{eskildsen},
the $\pi$-gap  states are filled  much more slowly  than in the
low field region. A magnitude of  the upper critical fields of
two  bands   can  be  estimated   in  the  clean   limit  
\cite{bouquet}  as $H_{c2,\pi}   \approx  \Delta_{\pi}^2/v_{F,
\pi}^2$  and $H_{c2,\sigma}  \approx \Delta_{\sigma}^2/v_{F,
\sigma}^2$, with $v_{F,\pi}$ or $v_{F,\sigma}$ for  the  Fermi velocity of the respective band
(since  only  $H_{c2||c}$ is considered the anisotropy due to an effective mass
tensor is neglected).
A similar
estimates can  be done in  the dirty limit  when taking into
account the respecive diffussion coefficients $D_{\pi(\sigma)}: H_{c2,\pi(\sigma)}
\approx \Delta_{\pi(\sigma)}/D_{\pi(\sigma)}$.

\begin{figure}[t]
\hspace{-8 mm}
\includegraphics[width= 9.4 cm ]{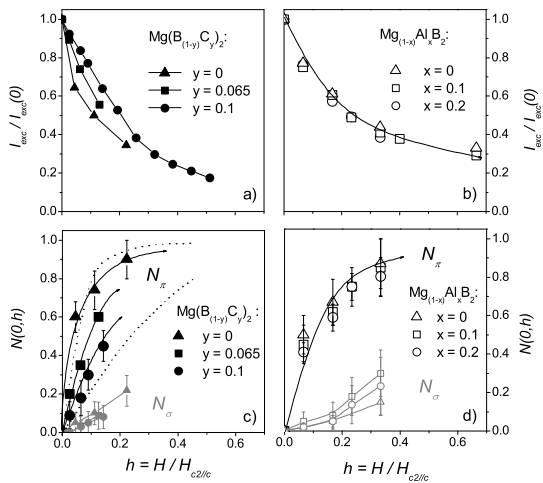}
\caption{a) Excess currents $I_{exc}(h)$ of Mg(B$_{(1-y)}$C$_y$)$_2$. b)  $I_{exc}(h)$ of Mg$_{1-x}$Al$_x$B$_2$. c) Zero-bias density of states $N(0,h)$ of Mg(B$_{(1-y)}$C$_y$)$_2$. d) Zero-bias DOS of Mg$_{1-x}$Al$_x$B$_2$ determined from the fitting for the two-band mixed state BTK formula. The solid lines are guides for the eyes. All here presented field dependencies have been determined from the point-contact spectra with similar weight of the $\pi$-band contribution $\alpha \approx 0.75\pm 0.05$. The dotted and dashed gray lines in Fig. 6c represents the theoretical predictions of Ref. \cite{koshelev} for the dirty Mg$B_{2}$ samples with the ratio of in-plane diffusivities $D_{\sigma} / D_{\pi} = 0.2$ and 1, respectively.}
\label{fig:fig6}
\end{figure}

Fig. 6a shows that the carbon doping leads to changes in the
field dependence of the excess current. The undoped MgB$_2$
reveals  a  very  steep  fall of $I_{exc}$  already at small fields
indicating also  proportionally small crossover  field $H_{c2,\pi}$. By doping the $\pi$-band crossover field apparently shifts
to higher values. From the  previous subsection we know that
the carbon  doping does not  influence significantly the interband scattering in MgB$_2$.
The  ratio between the two  gaps, $\Delta_{\sigma} / \Delta_{\pi}$ remains almost unchanged.
Then, following from the dirty limit theory the shift of the normalized crossover field $H_{c2,\pi} / H_{c2||c}$
is due to an increase of the ratio of    the    diffussion    coeficients
$D_{\sigma}/D_{\pi}$.  This is an indication that the
carbon  doping leads  to  a  significant enhancement  of the
$\pi$-band scattering.

In Fig. 6b the excess currents on the aluminum doped samples
show different picture: there is no change in the field
dependence of  $I_{exc}$ upon the  aluminum doping up  to 20
\% indicating constant value of the crossover field $H_{c2,\pi}$. 
Qualitatively, it can be  understood in the framework of
the clean limit persisting in  the Al-doped MgB$_2$ up to 20
\% Al doping region. Indeed, the constant value of the upper
critical field $H_{c2||c}$, which  is equal to approximately
3 Tesla for all three  dopings, indicates the constant ratio
between  $\Delta_{\sigma}$  and  $v_{F,\sigma}$. It means
that  the   electron  filling  of   the  hole  $\sigma$-band
decreases  both  quantities  proportionally.  Then,  $H_{c2,
\pi}$ must  be only weakly doping  dependent. As $H_{c2,\pi}
\approx  \Delta_{\pi}^2/v_{F,   \pi}^2$  and  $\Delta_{\pi}$
remains  almost constant  for x   = 0.1  and 0.2  this would
suggest  that  $v_{F,  \pi}$  remains  also  unchanged.

Quantitatively, $I_{exc}$  of the two-gap  superconductor in
the mixed  state can be described  by the sum \cite{naidyuk}
\begin{equation}        I_{exc}        \approx        \alpha
(1-n_{\pi})\Delta_{\pi}                                    +
(1-\alpha)(1-n_{\sigma})\Delta_{\sigma}.  \end{equation} 

For
the  point-contact conductance  a similar  expression can be
written  \begin{equation}   G  \approx  \alpha   [n_{\pi}  +
(1-n_{\pi})]G_{\pi}                                        +
(1-\alpha)[n_{\sigma}+(1-n_{\sigma})]G_{\sigma}.
\end{equation}

Considering  that  $n_{\pi}$  and  $n_{\sigma}$ represent independent  fillings  of  the  respective  gaps in increasing
applied  field,  their  values  can  be  identified with the
values of the zero-energy density  of states (DOS) $N_{\pi}(0,H)$  and
$N_{\sigma}(0,H)$ averaged over the vortex lattice.    This   
model    proposed    by Bugoslavsky {\it et al.}
\cite{bugoslavsky}    enables    a    direct    experimental
determination of  the field dependent energy gaps
and zero-energy DOS $N(0,H)$ from the point-contact spectra in
MgB$_2$. The calculations of Koshelev and Golubov \cite{koshelev} show that the low field slope of $N_{\pi}(0,H)$ is controlled by  an important parameter $D_{\sigma}/D_{\pi}$ which can thus be established.

 Unfortunately the fitting of the PC conductance of MgB$_2$ in the mixed state is not trivial because of
too many (9) parameters. Moreover two
pairs of  the parameters $N_{\pi}$  vs. $\Gamma _{\pi}$  and
$N_{\sigma}$ vs. $\Gamma  _{\sigma}$ manifest very similarly
in  the  fit.  Hence,  before  fitting  the spectra by the formula (3) we  
estimate $N_{\pi}(0,H)$ values  from  the excess currents.
Here some simplifications were made: 
If the real upper critical field $H_{c2}$ is large compared to $H_{c2,\pi}$, in
the region $0 < H < H_{c2,\pi}$ the suppression of
the $\sigma$ band contribution by the field can be neglected
($N_{\sigma}(0,H)  =  0)$.  If   we  also  neglect  a  small
reduction   of    the   energy   gaps    at   these   fields
\cite{bugoslavsky,samkreta},   then   $I_{exc}(H)$  can be represented by  a single parameter  formula 
expressed   as   $I_{exc}   \propto   \alpha   (1-N_{\pi}(0,
H))\Delta_{\pi}(0)+    (1-\alpha)\Delta_{\sigma}(0)$   
since all other 
parameters  ($\Delta_{\pi}(0)$,  $\Delta_{\sigma}(0)$ and $\alpha$) have
already been  determined from the BTK  fit at $H =  0$. 
The resulting 
$N_{\pi}(0,H)$ is later used as a first approximation in
the fit of the PC conductance spectrum. There, also
the values of $z_{\pi}, z_{\sigma},
\Gamma_{\pi}$ and $\Gamma_{\sigma}$ determined from BTK fits
at  $H  =  0$  are kept unchanged.
In the  second step all parameters except for  $z_{\pi}, z_{\sigma}$ and $\alpha$ are adjusted for the best fit.
The resulting  values  of $N_{\pi}(0,H)$
are decreased  by  5 - 20\% in comparison with the first estimate from  $I_{exc}(H)$ \cite{szabo05}.
Also the values  of   the  smearing  parameters   $\Gamma_{\pi}$  and
$\Gamma_{\sigma}$  reveal about $10\%-20\%$  of increase
in the  interval $0 < H  < 1$ T accounting for the magnetic pair-breaking of the superconducing DOS. In  the same field interval a  small  $\approx (5-10\%$) suppression of  the values of
the energy  gaps has been  obtained. 

The resulting  zero-energy DOS $N_{\pi}(0,H)$ and $N_{\sigma}(0,H)$ are
plotted in Fig.6c for the carbon-doped samples and in Fig.6d for the Al-substituted MgB$_2$.
Again, the carbon and aluminium dopings reveal very differently.
The steep  increase of  $N_{\pi}(0,h)$ in  the undoped MgB$_2$ can   be  ascribed  within   the  framework  of the  dirty
superconductivity model \cite{koshelev} to  a small value of the ratio  of in-plane diffusivities  $D_{\sigma} /D_{\pi} < 1$ (see dotted line, calculated for $D_{\sigma} /D_{\pi} = 0.2$). It means a dirtier $\sigma$-band than
the $\pi$-one.  The subsequent decrease  of the
low  field  slope  of  $N_{\pi}(0,h)$  upon  C-doping can be
explained  by an increase of  the  $D_{\sigma} /D_{\pi}$ ratio, i.e. by a 
more rapid enhancement of the $\pi$-band  intraband scattering as compared with the scattering in the $\sigma$-band.
The dashed line plots the theoretical $N_{\pi}(0,h)$ dependence at $D_{\sigma} /D_{\pi} = 1$ 

The situation is  different for the Al-doping. Here the
field dependences  of $N_{\pi}(0,h)$ 
reveal no change upon doping.  The same explanation as used above for the excess current versus  field can be applied, underlying the conservation of the superconducting clean limit for Al doping.

\section{CONCLUSIONS}

The influence of carbon and aluminium doping on the two-band/two-gap superconductivity in Mg$B_2$ has been studied by the point-contact spectroscopy. It has been shown, that the decreased transition temperatures and the evolution of the gaps upon both substitutions is mainly a consequence of the band filling effect. However, in the case of an increased Al-doping also an increase of the interband scattering has to be taken into account. 
By the analysis of the field dependences of the point-contact spectra we have shown, that the C-doping increases the scattering in the $\pi$-band more rapidly then in the $\sigma$ one. On the other hand, the Al-doping does not introduce significant changes in the relative weight of the scatterings within two bands.

\acknowledgments
This  work  has  been  supported  by  the Slovak Science and
Technology     Assistance   Agency     under     contract
No.APVT-51-016604.  Centre of Low Temperature Physics is
operated as  the Centre of Excellence  of the Slovak Academy
of Sciences under contract  no. I/2/2003. Ames Laboratory is
operated  for the  U.S. Department  of Energy  by Iowa State
University under  Contract No. W-7405-Eng-82.  This work was
supported  by the  Director for  Energy Research,  Office of
Basic   Energy  Sciences.   The  liquid   nitrogen  for  the
experiment has  been sponsored by the  U.S. Steel Ko\v sice,
s.r.o.


\begin{thebibliography}{99}
\bibitem{footnote}
Due to measurements on polycrystals the field orientation is not fully controlled. But in the present study we are interested mainly in the behavior of the isotropic $\pi$ band, thus the choice of the crystallographic orientation of the "reference" upper critical field is arbitrary but once chosen it must be kept for all compared doped samples.
\end{thebibliography}
\end{document}